\documentclass[aps,pre,reprint,superscriptaddress,amsmath,amssymb,showpacs,floatfix,longbibliography]{revtex4-1}
\usepackage{graphicx}
\usepackage{bm}
\usepackage{color}

\newcommand{\RN}[1]{\textup{\uppercase\expandafter{\romannumeral#1}}}
\newcommand{\bmf}{{\bm{f}}}
\newcommand{\bmx}{{\bm{x}}}
\newcommand{\bmy}{{\bm{y}}}
\newcommand{\bmz}{{\hat{\bm{x}}}}

\newcommand{\bmxi}{{\bm{\xi}}}
\newcommand{\bmzeta}{{\bm{\zeta}}}

\def\dbar{{\mathchar'26\mkern-12mu d}}

\begin{document}

\title{Efficiency at maximum power and efficiency fluctuations 
in a linear Brownian heat engine model} 

\author{Jong-Min Park}
\affiliation{Department of Physics, University of Seoul, Seoul 02504,
Korea}
\author{Hyun-Myung Chun}
\affiliation{Department of Physics, University of Seoul, Seoul 02504,
Korea}
\author{Jae Dong Noh}
\affiliation{Department of Physics, University of Seoul, Seoul 02504,
Korea}
\affiliation{School of Physics, Korea Institute for Advanced Study,
Seoul 02455, Korea}

\date{\today}

\begin{abstract}
We investigate stochastic thermodynamics of a two-particles Langevin system.
Each particle is in contact with a heat bath at different
temperatures $T_1$ and $T_2~(<T_1)$, respectively. 
Particles are trapped by a harmonic potential and driven by a linear external 
force. 
The system can act as an autonomous heat engine performing work against the 
external driving force. Linearity of the system enables us to examine
thermodynamic properties of the engine analytically.  
We find that the efficiency of the engine at maximum power $\eta_{MP}$
is given by $\eta_{MP} = 1-\sqrt{T_2/T_1}$. 
This universal form has been known as a 
characteristic of endoreversible heat engines. Our result extends the
universal behavior of $\eta_{MP}$ to non-endoreversible engines.
We also obtain the large deviation function of the probability
distribution for the stochastic efficiency 
in the overdamped limit. The large deviation function takes the minimum
value at mean efficiency $\eta = \bar{\eta}$ and increases monotonically 
until it reaches plateaus when $\eta \leq \eta_L$ and $\eta \geq
\eta_R$ with model dependent parameters $\eta_{R,L}$. 
It has been known for heat engines with a finite number of
microscopic configurations with time-symmetric protocol 
that the probability of achieving the Carnot efficiency is minimum. 
Our result reveals that the least likeliness of the Carnot 
efficiency is not the generic property of heat engines.

\end{abstract}
\maketitle

\section{Introduction}

Heat engines are devices to generate mechanical work 
by exploiting heat flows between hot and cold heat baths at temperatures 
$T_1$ and $T_2~(<T_1)$. 
Since the advance of stochastic thermodynamics, Brownian heat engines
consisting of microscopic small components have been attracting a lot of
theoretical and experimental interests. Those engines are working in
nonequilibrium conditions and subject to large thermal fluctuations. 
Much efforts have been devoted to understanding common properties that
are shared by a variety of different engine models.

The efficiency $\eta$, defined as the ratio of the work to the absorbed 
heat from a hot heat bath, is one of the most important characteristics of
a heat engine. 
According to the thermodynamic laws, the efficiency is limited from above 
by the Carnot efficiency $\eta_{C} \equiv 1-T_2/T_1$. 
The Carnot efficiency is achieved only when an engine operates 
infinitely slow and reversibly. Hence, an engine operating at the Carnot
efficiency is of no practical importance because its power, work per unit
time, is zero. 

Instead of optimizing the efficiency, researchers are interested in the 
efficiency of an engine when it is optimized to yield the maximum power, 
which is called the efficiency at maximum power~(EMP) $\eta_{MP}$. 
The EMP is shown to be universal for endoreversible engines that
operate reversibly except when they exchange heats with external 
heat baths~\cite{Curzon:1975gt}. 
The EMP of the endoreversible engines is given by 
$\eta_{MP} = \eta_{CA} \equiv 1-\sqrt{T_2/T_1}$. This efficiency $\eta_{CA}$
is called the Curzon-Ahlborn efficiency since it was rediscovered by 
Curzon and Ahlborn~\cite{Curzon:1975gt} while it was first known long
before~\cite{Chambadal:1957aa,Novikov:1958aa}. 

Most of realistic engines are not
endoreversible~\cite{Sekulic:1998aa,Andersen:2001aa}. 
Nevertheless, the EMP of many engines is close to $\eta_{CA}$ 
when $T_1$ and $T_2$ are close to each other
so that $\eta_C \ll 1$. In this limit, the Curzon-Ahlborn efficiency is
expanded as $\eta_{CA} = 1-\sqrt{1-\eta_C} = \frac{1}{2}\eta_C +
\frac{1}{8}\eta_C^2 +O(\eta_C^3)$. Some engines, which are not
endoreversible, share the same expansion up
to first or second order in
$\eta_C$~\cite{Gomez-Marin:2006aa,Schmiedl:2008aa,Tu:2008aa,Izumida:2008aa,
Izumida:2009aa,Izumida:2009ab,Esposito:2009aa,Esposito:2010aa,
Esposito:2012aa,Apertet:2012aa,Apertet:2012ab,Van-den-Broeck:2012aa,
Schmiedl:2008ab,Zhou:2010aa,Seifert:2011aa,Golubeva:2012aa,
Van-den-Broeck:2012ab,Golubeva:2012ab,Golubeva:2013aa,Hooyberghs:2013aa,
Golubeva:2014aa,Allahverdyan:2008aa,Abe:2011if}. 
It was found that the first 
order term reflects the strong coupling between thermodynamic
fluxes~\cite{Van-den-Broeck:2005aa} and that the second order term the
left-right symmetry~\cite{Esposito:2009ke}. 
The universality of the expansion has been investigated in the context of
irreversible thermodynamics~\cite{Cleuren:2015aa}. 

When one measures the efficiency of an engine for a time interval
$t$, it varies from one measurement to another due to thermal
fluctuations. Thus, the efficiency is a fluctuating random variable 
characterized by the probability distribution function
$P_t(\eta)$ and the large deviation function 
$L(\eta) \equiv -\lim_{t\to\infty}\frac{1}{t}\ln P_t(\eta)$
in the long time limit.
Recently, it was found that the large deviation function $L(\eta)$ is 
maximum at $\eta=\eta_C$. This means that the Carnot efficiency is least likely
in the $t\to\infty$ limit. To be precise, such a property was proved for
a heat engine which has only a finite number of microscopic configurations 
and is driven by a time-symmetric protocol~\cite{Verley:2014ha,Verley:2014hq}.
The least likeliness of the Carnot efficiency was demonstrated in two-level
systems analytically and numerically. However, it remains as an open 
question whether it is valid for systems with continuous variables.

In this paper, we introduce an exactly solvable model for a Brownian heat
engine. The model consists of two Brownian particles in one dimension which
are trapped by a harmonic potential and driven by a linear external force.
Each particle is in contact with a heat bath at different temperatures. 
The temperature difference induces a heat flow, which enables the system to
work against the external force. Owing to solvability, the linear systems 
have been adopted for detailed study of various subjects in stochastic
thermodynamics such as the entropy production, the fluctuation theorems,
information engines, and so on~\cite{Kwon:2005uq,Kwon:2011fk,Noh:2013jm,Chun:2015aa}. 
We will investigate thoroughly the linear model in the perspective of the
heat engine with the focus on the efficiency of the heat engine. 
Our results can be summarized as follows: (i) The exact expressions for the
average efficiency and power are derived.  We find that the EMP is equal to
$\eta_{CA}$. Our engine model operates in a nonequilibrium condition, hence
is not an endoreversible engine. This result indicates that the 
endoreversibility is not a necessary condition for $\eta_{MP} = \eta_{CA}$.
(ii) The large deviation function $L(\eta)$ for the efficiency is 
obtained analytically. The function is minimum at the average
efficiency, increases monotonically as $\eta$ departs from
the average efficiency, and reaches constant plateaus in the regions with 
$\eta \geq \eta_R$ and $\eta \leq \eta_L$. 
The large deviation function does not have a peak at the Carnot efficiency, 
which is in sharp contrast to the property of
finite-configurations heat engines.

This paper is organized as follows. We introduce the model system 
and calculate the steady state average of the heat
and work in Sec.~\ref{sec:Model}. 
We elaborate on the EMP and compare it with $\eta_{CA}$ in Sec.~\ref{sec:EMP}. 
In Sec.~\ref{sec:etaLDF}, we derive the exact expression for the large
deviation function for the efficiency. We summarize our results in
Sec.~\ref{sec:Conclusion}

\section{Linear engine model}\label{sec:Model}

We consider a system consisting of two Brownian particles of mass $m$ 
in one dimension. Two particles are in contact with two
different heat baths at temperatures $T_1$ and $T_2~(<T_1)$, respectively,
and linear forces are applied.
Their motions are governed by the underdamped Langevin equations
\begin{equation}\label{eq:LangevinEq}
\begin{aligned}
\dot{x}_1 &= v_1,\\
\dot{x}_2 &= v_2,\\
m \dot{v}_1 &= -\gamma v_1 - K x_1 + \epsilon x_2 + \xi_1(t) , \\ 
m \dot{v}_2 &= -\gamma v_2 - K x_2 + \delta x_1 + \xi_2(t) ,
\end{aligned}
\end{equation}
where $x_i$ and $v_i$ are the position and the velocity of $i(=1,2)$th 
particle, $\gamma$ is a damping coefficient, $K$ is a stiffness constant 
of a harmonic
potential trapping the particles at the origin, $(\epsilon,\delta)$ are the coupling constants, and $\xi_i(t)$
is the Gaussian-distributed random force satisfying $\langle
\xi_i(t)\rangle=0$ and $\langle \xi_i(t)\xi_j(t') \rangle = 2\gamma k_{B}
T_i \delta_{ij} \delta(t-t')$. 
We use a shorthand notation $\dot{}$ for a time derivative and 
set the Boltzmann constant $k_B$ to be unity hereafter.

The two-particle system may be interpreted as a single Brownian particle
system in two dimensions with position column vector $\bm{x}=(x_1,x_2)^T$ and
velocity column vector $\bm{v}=(v_1,v_2)^T=\dot{\bmx}$. The superscript
${}^T$ stands for the transpose.
In this interpretation, the total applied force $\bmf$ is decomposed into
the sum of two parts: $\bmf = \bmf_c+\bmf_{nc}$ with 
the conservative force 
\begin{equation}
\bm{f}_{{c}}=-K\bm{x} = -\bm{\nabla} V(\bm{x})
\end{equation} 
with a harmonic potential $V(\bm{x}) = \frac{1}{2}K\bm{x}^2$
and the nonconservative driving force 
\begin{equation}
\bm{f}_{{nc}}=(\epsilon x_2, \delta x_1)^T
\end{equation}  which
does not have a corresponding potential function 
unless $\epsilon = \delta$. The motions along the $x_1$-axis and the 
$x_2$-axis are affected independently by the heat baths of temperatures $T_1$ 
and $T_2$, respectively. 

For appropriate choices of $\epsilon$ and $\delta$, the system can work
against the nonconservative force by exploiting the heat flow between the heat
baths. Thus it can act as a heat engine as well as a heat pump or 
a refrigerator. 
According to stochastic energetics~\cite{Sekimoto:1998uf}, the heats absorbed
from the heat baths into the system and the work done by the system against
the driving force during an infinitesimal time interval 
$[t, t+dt]$ are given by
\begin{equation}\label{eq:HeatWork}
\begin{aligned}
\dbar Q_1(t) &= v_1(t) \circ \big[ -\gamma v_1(t) dt +d\Xi_1(t) \big], \\
\dbar Q_2(t) &= v_2(t) \circ \big[ -\gamma v_2(t) dt +d\Xi_2(t) \big],\\
\dbar W(t) &= -\bm{f}_{ nc}\circ d\bm{x} = -\big[ 
\epsilon v_1(t) x_2(t) + \delta x_1(t) v_2(t) \big] dt,
\end{aligned}
\end{equation}
where $d\Xi_i(t) \equiv \int_t^{t+dt} dt' \xi_i(t')$ are Gaussian random
variables satisfying $\langle d\Xi_i(t) \rangle = 0$ and $\langle d\Xi_i(t)
d\Xi_j(t) \rangle = 2 \gamma T_i \delta_{ij} dt$. 
The notation $\circ$ represents the Stratonovich 
product~\cite{Risken:1996vl,Gardiner:2010tp}. Those quantities satisfy
the energy conservation $dE(t)=\dbar Q_1(t)+\dbar Q_2(t)-\dbar W(t)$ with
the internal energy $E = \frac{1}{2}m\bm{v}^2 + V(\bmx)$.

We focus on the average quantities in the steady state, denoted by $\langle
\cdot \rangle_s$. Fluctuations are considered later. 
The steady-state average of the internal 
energy change, $\langle dE\rangle_s$, vanishes. Hence,
there exist only two relevant quantities describing the energy flow. We
choose the heat flow rate from the hot reservoir 
$q_1 \equiv \langle \dbar Q_1/dt\rangle$
and the work production rate $w \equiv \langle \dbar W/dt\rangle_s$.
The Stratonovich algebra yields that $\langle v_1(t) \circ
d\Xi_1(t)\rangle_s = \langle \frac{1}{2}(v_1(t)+v_1(t+dt)) d\Xi_1(t)
\rangle_s = \frac{\gamma T_1}{m}dt + o(dt)$. Thus, we obtain the
expressions~\cite{Rieder:1967fi,Tome:2010gd,Lippiello:2014jf,Chun:2015aa}
\begin{equation}\label{eq:HeatWorkAv}
\begin{aligned}
q_1 &= \frac{2\gamma}{m} \left( \frac{T_1}{2} - 
\frac{1}{2} m \left\langle v_1^2 \right\rangle_s \right) ,\\
w &= -\epsilon \left\langle v_1 x_2 \right\rangle_s - 
\delta \left\langle x_1 v_2 \right\rangle_s.
\end{aligned}
\end{equation}
The average heat flux from the cold reservoir is
given by $q_2 \equiv \langle \dbar Q_2/dt\rangle_s = w-q_1$.
%

The Langevin equations in \eqref{eq:LangevinEq} are linear in $\bm{z} = (x_1,x_2,v_1,v_2)^T$
and belong to the class of the multivariate Ornstein-Uhlenbeck
process~\cite{Risken:1996vl,Gardiner:2010tp}. 
In such a case, the steady state is Gaussian-distributed with the covariance
matrix $\mathsf{\Sigma} = \langle \bm{z} \bm{z}^T \rangle_s$ being
determined as a solution of a set of linear equations. Following the standard
procedure~(see Sec.4.5.6 of Ref.~\cite{Gardiner:2010tp}), we obtain that
\begin{equation}\label{eq:2ndMoments}
\mathsf{\Sigma} = \begin{pmatrix}
\frac{(K\psi + \gamma^2 \phi)}{\delta} & \psi & 0 & \gamma \phi \\
\psi & \frac{( K\psi - \gamma^2 \phi)}{\epsilon} & -\gamma \phi & 0 \\
0 & -\gamma \phi & \frac{T_1}{m} - \epsilon \phi & 0 \\
\gamma \phi & 0 & 0  & \frac{T_2}{m} + \delta \phi
\end{pmatrix}
\end{equation}
with $\psi=\frac{\delta T_1 + \epsilon T_2}{2(K^2-\epsilon \delta)}$ and
$\phi=\frac{\delta T_1 - \epsilon T_2}{2(\gamma^2 K +m\epsilon \delta)}$.
Using the covariance matrix, we find that
\begin{equation}\label{eq:HeatWorkFin}
\begin{aligned}
q_1 &= \gamma \epsilon \phi = \frac{\gamma \epsilon (\delta T_1 - \epsilon T_2)}{2(\gamma^2 K + m\epsilon \delta)}\\
w&= \gamma (\epsilon-\delta) \phi = \frac{\gamma (\epsilon-\delta) (\delta T_1 - \epsilon T_2)}{2(\gamma^2 K + m\epsilon \delta)}.
\end{aligned}
\end{equation}
The covariance matrix is positive-definite in the region 
\begin{equation}\label{stable}
-\frac{\gamma^2 K}{m} < \epsilon \delta < K^2 .
\end{equation}
Outside the region, the nonconservative force is so strong that the particle
escapes from the harmonic potential. We will restrict ourselves to the
stable region for further analysis.

\section{Efficiency at maximum power}\label{sec:EMP}

\begin{figure}
\includegraphics*[width=0.8\columnwidth]{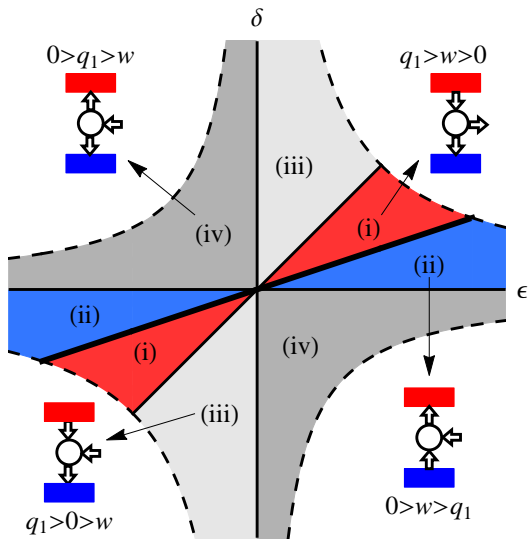}
\caption{Function diagram of the linear engine model.
The dashed lines are the boundary of the stable region.
}\label{fig1}
\end{figure}

As one varies $\epsilon$ and $\delta$ within the stable region, $q_1$ and $w$
flip their signs. There are four different regions:
(i)~When $q_1>0$ and $w>0$, the system operates as a heat engine
which absorbs a heat from the hot bath, dissipates a heat to a cold bath, 
and works against the driving force.
The average engine efficiency is given by
\begin{equation}\label{eq:eta}
\bar\eta = \frac{w}{q_1} = 1 - \frac{\delta}{\epsilon} .
\end{equation}
We use the notation $\bar\eta$ for the average efficiency 
in order to distinguish it from the stochastic efficiency $\eta$ 
investigated later.
(ii)~When $q_1<0$, $w<0$ and $q_2 = w-q_1>0$, the system operates
as a heat pump or a refrigerator which transfers a heat from the cold
bath~($q_2$) to the hot bath~($|q_1|$) with the help of an external
work~($|w|$). 
(iii)~When $q_1>0$, $w<0$, and $q_2<0$, a heat flows from the hot bath to the
cold bath at the expense of an external work. 
(iv)~When $q_1<0$, $w<0$ and $q_2<0$, an external work is dissipated into the
two baths. The border lines of these four regions and the stable region are
drawn in Fig.~\ref{fig1}.
The two regions (iii) and (iv) are of no practical importance. 
We focus on the heat engine region (i).

The regions (i) and (iii) are separated by the line
$\delta=\epsilon$, where the force $\bm{f}_{nc}$ becomes
a conservative one. Hence the power $w$ vanishes and the system plays a role
of a heat conductor. 

The heat engine regime (i) is separated from the heat pump regime (ii) by the 
line $\delta = (T_2/T_1)\epsilon = (1-\eta_C) \epsilon$, drawn with the
thick line in Fig.~\ref{fig1}. Along this line,
the efficiency in \eqref{eq:eta} is given by the Carnot efficiency $\eta_C =
1-T_2/T_1$ with vanishing power~(see \eqref{eq:HeatWorkFin}).
In macroscopic thermodynamics, the Carnot efficiency is achieved only
when an engine operates quasi-statically and reversibly. 
The vanishing power and the Carnot efficiency along the line are thus 
consistent with each
other~\cite{Gomez-Marin:2006aa,Tu:2008aa,Esposito:2009aa}.
In fact, our model can be shown to be in thermal equilibrium along the line 
$\delta = (1-\eta_C)\epsilon$. 
In terms of dimensionless parameters 
$\tilde{x}_1=\frac{x_1}{(\sqrt{mT_1}/\gamma)}$,
$\tilde{x}_2=\frac{x_2}{(\sqrt{mT_2}/\gamma)}$, and
$\tilde{t}=\frac{t}{(m/\gamma)}$, the Langevin equation
\eqref{eq:LangevinEq} becomes equivalent to that for 
a two-dimensional Brownian particle in thermal contact with a single
heat bath at unit temperature. The particle is
driven by the effective nonconservative force
$\tilde{\bm{f}}_{{nc}}=\left(\frac{m}{\gamma^2} \sqrt{\frac{T_2}{T_1}}
\epsilon \tilde{x}_2, \frac{m}{\gamma^2} \sqrt{\frac{T_1}{T_2}} \delta
\tilde{x}_1\right)^T$, which turns into the conservative force along the
line $\delta=(T_2/T_1)\epsilon = (1-\eta_C)\epsilon$. 
The temperature difference~($T_1\neq T_2$)
and the nonconservative force $\bm{f}_{\rm nc}$ are the ingredients that
drive the system out of equilibrium.
When $\delta T_1=\epsilon T_2$, their effects cancel each other and 
the system is in thermal equilibrium.

\begin{figure}
\includegraphics*[width=0.9\columnwidth]{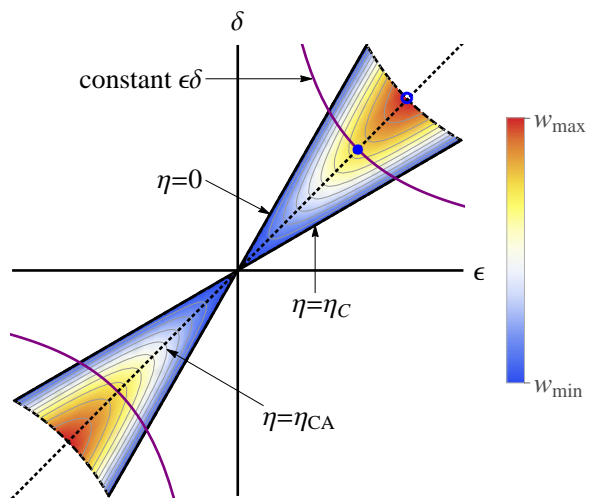}
\caption{Density and contour plots for the engine power $w$.
}\label{fig2}
\end{figure}
The power $w$ of the engine varies in the $(\epsilon,\delta)$ plane as shown
in Fig.~\ref{fig2}. We will find the maximum power point and 
investigate how the EMP depends on the temperatures.
The power $w$ is given by a function of $\epsilon$ and $\delta$ 
in \eqref{eq:HeatWorkFin}. Recalling that the average efficiency $\bar\eta$ in
\eqref{eq:eta} is a function of $\delta/\epsilon$, we found it convenient to
write $w$ as a function of $\epsilon \delta$ and
$\bar\eta$ instead of a function of $\epsilon$ and $\delta$:
\begin{equation}\label{eq:Power}
w(\epsilon\delta,\bar\eta)=\frac{\gamma \epsilon \delta T_1}{2(\gamma^2 K+m
\epsilon \delta)} \frac{\bar\eta(\eta_{{C}}-\bar\eta)}{1-\bar\eta}
\end{equation}
with $\eta_C = 1-T_2/T_1$.
Then, for a given $\epsilon \delta$, the power is maximum when 
$\frac{\partial w}{\partial \bar\eta}=0$, which
yields that
\begin{equation}\label{eq:eta_emp}
\eta_{MP} = 1-\sqrt{1-\eta_C} \ .
\end{equation}
This is the EMP along the constant-$\epsilon \delta$ 
curves~(see Fig.~\ref{fig2}). The global maximum of the
power is achieved in the limiting case where $\epsilon\delta$ approaches 
$K^2$, the border of the stable region~(see \eqref{stable}), the efficiency
at which is also given by \eqref{eq:eta_emp}.

To our surprise, the result for the efficiency at maximum power is the same
as the Curzon-Ahlborn efficiency $\eta_{CA}$ obtained for 
the endoreversible engine~\cite{Curzon:1975gt}. It reveals that the
endoreversibility is not a necessary condition for $\eta_{MP} =
\eta_{CA}$. In order to understand the similarity between our model and the
endoreversible engines, we rederive the Curzon-Ahlborn
result~\cite{Curzon:1975gt,De-Vos:1985aa,Bejan:1988aa}.
An endoreversible engine operates under the assumption that it maintains
internal temperatures $T_{1i}$ and $T_{2i}$ 
when it exchanges heats with the heat baths at temperatures at $T_1$ and 
$T_2$, respectively. The endoreversibility means that the engine 
operates as the Carnot engine between two temperatures $T_{1i}$ and
$T_{2i}$.
Assuming the Fourier law, the incoming~($q_1$) and outgoing~($-q_2$) 
heat fluxes are 
given by $q_1 = \alpha_1 (T_1 - T_{1i})$ and $-q_2 = \alpha_2 (T_{2i}-T_2)$,
respectively, with the heat conductivities $\alpha_i$. 
Then, the endoreversible condition amounts to $q_1/T_{1i} = -q_2/T_{2i}$.
The power is given by $w = q_1 + q_2 = \alpha_1 
(T_1-T_{1i}) - \alpha_2 (T_{2i}-T_2)$.
It is a function of the internal temperatures $T_{1i}$ and $T_{2i}$
which are determined by an operating condition.
Using the endoreversible condition and the expression for the efficiency 
$\bar\eta = 1-\frac{\alpha_2(T_{2i}-T_2)}{\alpha_1 (T_1-T_{1i})}$, 
one can eliminate $T_{1i}$ and $T_{2i}$ in $w$ to obtain that 
\begin{equation}\label{eq:CA_power}
w = \frac{\alpha_1 \alpha_2 T_1}{\alpha_1+\alpha_2} 
\frac{\bar\eta(\eta_{{C}}-\bar\eta)}{1-\bar\eta}
\end{equation}
Apart from the overall factor, it has the same $\bar\eta$-dependence as in
\eqref{eq:Power}, hence the same efficiency at maximum power.

Comparing \eqref{eq:Power} and \eqref{eq:CA_power}, one finds that the
Curzon-Ahlborn efficiency $\eta_{CA} = 1-\sqrt{T_2/T_1}$ is the consequence
of the specific relation between the thermodynamic quantities irrespective
of microscopic details of engines. 
It is convenient to use the parameters $s_1 = q_1/T_1$~(entropy loss of the
hot bath) and $s_2 = -q_2 /T_2$~(entropy gain of the cold bath). 
Using $w = s_1 T_1 - s_2 T_2$ and $\bar\eta = 1-(s_2 T_2)/(s_1
T_1)$, we can rewrite \eqref{eq:Power} and \eqref{eq:CA_power} in the form
$s_1 = \mathcal{F}(s_2)$ where the function $\mathcal{F}(x)$ is given by
\begin{equation}\label{eq:f(x)}
\mathcal{F}(x) = \frac{x}{1+\zeta x}
\end{equation}
with $\zeta = 2(\gamma^2 K+m \epsilon\delta)/(\gamma \epsilon \delta)$ for
\eqref{eq:Power} and $\zeta=\alpha_1^{-1} + \alpha_2^{-1}$ for
\eqref{eq:CA_power}.

\begin{figure}
\includegraphics*[width=0.9\columnwidth]{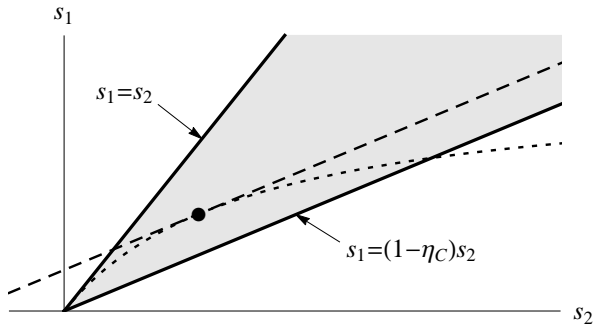}
\caption{Linear heat engine model in the $(s_1,s_2)$ plane. 
The model acts as a heat engine in the shaded area satisfying 
$(1-\eta_C ) s_2 < s_1 < s_2$. The dotted curve represents a characteristic
of an engine. The maximum power is achieved when the curve is tangential to
a straight line of slope $(1-\eta_C)$.
}\label{fig3}
\end{figure}

In general, as one varies engine-specific parameters, 
such as $\epsilon$ and $\delta$ in
our model or $T_{1i}$ and $T_{2i}$ in the endoreversible engine, $s_1$ and
$s_2$ will move along a curve $s_1 = \mathcal{F}(s_2)$. 
We now address the question 
whether the function $\mathcal{F}(x)$ in \eqref{eq:f(x)} is uniquely determined 
for all systems displaying the Curzon-Ahlborn efficiency. 
In Fig.~\ref{fig3}, we draw an arbitrary curve~(dotted line) in $(s_1,s_2)$ 
plane.
The thermodynamic second law $s_1 \leq s_2$ requires that  
the function $\mathcal{F}(x)$ should be below the straight line $s_1=s_2$. 
The device works as a heat engine when $s_1>0$, $s_2>0$, and 
$w = q_1+q_2  = T_1 s_1 - T_2 s_2\geq 0$. Hence, the shaded area between two
straight lines $s_1 = s_2$ and $s_1 = (T_2/T_1)s_2 = (1-\eta_C) s_2$ 
is the region of physical interest. Noting that the power of the engine 
is constant along a straight line $s_1 = (1-\eta_C) s_2 + w/T_1$, one finds
that the maximum power achieved when the curve $s_1 = \mathcal{F}(s_2)$ is tangential
with the straight line of slope $(1-\eta_C)$. The tangential point ($s_1^*,
s_2^*)$, hence the maximum power point, is determined by 
\begin{equation}\label{de1}
s_1^* = \mathcal{F}(s_2^*) \ , \ (1-\eta_C) = \mathcal{F}'(s_2^*) \ .
\end{equation}
The efficiency at maximum power is then given by
\begin{equation}\label{de2}
\eta_{MP} = 1 -(1-\eta_C) \frac{s_2^*}{s_1^*} \ .
\end{equation}
We now impose that $\eta_{MP} = 1-\sqrt{1-\eta_C}$ for any combinations
for $T_1$ and $T_2$, i.e., any value of $\eta_C$. Eliminating $\eta_C$ using
\eqref{de1} and \eqref{de2}, we obtain the differential equation for the
function $\mathcal{F}(x)$:
\begin{equation}
\mathcal{F}'(x) = \frac{\mathcal{F}(x)^2}{x^2} \ .
\end{equation}
The solution of the differential equation is given by the function 
in \eqref{eq:f(x)}. This analysis shows that the Curzon-Ahlborn efficiency
at maximum power is achieved if and only if the entropy loss rate in the hot
reservoir and the entropy gain rate in the cold reservoir are constrained 
by the function given in \eqref{eq:f(x)}.

We add a few remarks. Firstly, there have been attempts to
understand the Curzon-Ahlborn efficiency from the symmetry consideration.
Near equilibrium where $T_1 \simeq T_2$ or $\eta_C = 1-T_2/T_1\ll 1$, the
Curzon-Ahlborn efficiency is expanded as $\eta_{CA} = \frac{1}{2}\eta_C +
\frac{1}{8}\eta_C^2 + O(\eta_C^3)$.
The first order term $\frac{1}{2}\eta_C$ reflects the strong coupling 
between thermodynamic fluxes~\cite{Van-den-Broeck:2005aa}. 
Namely, the heat fluxes and mechanical 
flux are proportional to each
other so that the total entropy production should be also proportional to
the heat flux or $s_1$. The function form $\mathcal{F}(x) = x/(1+\zeta x)$
implies that the total entropy production rate is given by $s_{tot} =
-s_1+s_2 = \zeta s_1 s_2$, which shows that our model belongs to the strong
coupling category. The second order term $\frac{1}{8}\eta_C^2$ is 
a manifestation of the so-called left-right symmetry~\cite{Esposito:2009ke}
under the exchange of the role between the hot and cold heat baths. 
Note that the relation $s_1 = \mathcal{F}(s_2)$ is invariant under the
changes $s_1 \to -s_2$ and $s_2 \to -s_1$ because the inverse of
$\mathcal{F}(x)$ is given by $\mathcal{F}^{-1}(x) = -\mathcal{F}(-x)$.
Thus, our model has the left-right symmetry. 
The higher order terms do not have a simple
explanation yet. Hopefully, our result may shed some lights in revealing the
physical meaning of the whole higher order terms.

Secondly, in general, as one varies microscopic parameters, 
an engine may cover the whole physical
region in the $(s_1,s_2)$  plane instead of following a one-dimensional
curve such as $s_1 = \mathcal{F}(s_2)$ in our model. 
Such a one-dimensional representation is possible when there
exist only a single independent parameter. The model of Curzon and
Ahlborn includes two parameters $T_{1i}$ and $T_{2i}$~\cite{{Curzon:1975gt}}. 
However, the endoreversibility condition eliminates one degree of freedom. 
In our model, we reduced the number of independent parameters by following 
the constant $\epsilon \delta$ curve. 
For general heat engines with multiple degrees of freedom,
if the entropy production rates of two reservoirs satisfy 
$s_1=\mathcal{F}(s_2)$ with a certain parameter $\zeta$, the efficiency 
at maximum power at constant $\zeta$ is always given by $\eta_{CA}$. 
It raises questions on the universality of $\eta_{CA}$ and on the role of
such a parameter $\zeta$, which are beyond the scope of the present study.

\section{Efficiency fluctuation}\label{sec:etaLDF}

The efficiency $\eta$ is a fluctuating random variable. 
Recent studies suggest that it is least probable that a nonequilibrium
heat engine would achieve the Carnot efficiency $\eta_C = 1-T_2/T_1$ 
in the long time limit~\cite{Verley:2014ha,Verley:2014hq}. This
result is derived for an engine which possesses a finite number of microscopic 
states and is driven by a time-symmetric protocol. 
Our engine is driven by a time-independent protocol which is obviously 
time-symmetric. However, its phase space is continuous with infinitely
many microscopic states. We will examine
whether the general statement of Refs.~\cite{Verley:2014ha,Verley:2014hq}
is also valid in our model.

For simplicity, we consider the overdamped dynamics.
Hereafter, the time will be rescaled so that the damping coefficient is 
taken to be unity.  
Then, the equations of motion for the position vector $\bmx = (x_1,x_2)^T$ 
are written as $\dot\bmx = \bmf + \bmxi$ with the force 
$\bmf = \bmf_{c} + \bmf_{nc}$ and the thermal noise 
$\bmxi = (\xi_1,\xi_2)^T$. 
Our task is to find the probability distribution $P_t(\eta)$ 
for the stochastic efficiency $\eta = W/Q_1$ where 
$Q_1$ is the heat absorbed from the hot reservoir and $W$ is the work
done against the nonconservative force $\bmf_{nc}$ up
to time $t$~(we will drop the subscript in $Q_1$ for notational convenience).
We focus on the large deviation function~(LDF)
\begin{equation}
L(\eta) \equiv -\lim_{t\to\infty} \frac{1}{t} \ln P_t(\eta) .
\end{equation}

In order to find $P_t(\eta)$, one needs to obtain 
the joint probability distribution $p(Q,W;t)$.
It is accessible by considering the Fokker-Planck equation for the probability 
distribution $p(\bmy;t)$ of the four-component vector 
$\bm{y} = (x_1,x_2, Q,W)^T$. This method was
introduced for the heat fluctuation of a one-dimensional Brownian
particle~\cite{Visco:2006en}. We extend the method to calculate the joint
distribution for $Q$ and $W$.
The generating function is defined as
\begin{equation}\label{G_gf}
G_t(x_1,x_2,\lambda_Q,\lambda_W)  \equiv \int dQ dW
e^{-\lambda_Q Q-\lambda_W W} p(\bmy;t) .
\end{equation}
After a lengthy algebra, we find that
\begin{equation}\label{Gt_sol}
G_t \propto \exp\left[ -\frac{1}{2} \bmx^T \cdot \mathsf{D}^{-1/2}
\mathsf{J} \mathsf{D}^{-1/2}\cdot \bmx + \mu(\lambda_Q,\lambda_W) t\right] 
\end{equation}
in the large $t$ limit, where
$\mathsf{J} = \mathsf{J}(\lambda_Q,\lambda_W)$ is a symmetric 
$2\times 2$ matrix and 
\begin{equation}\label{mu_res}
\mu(\lambda_Q,\lambda_W) = H(\lambda_Q + \bar\eta \lambda_W)
\end{equation}
with the function
\begin{equation}\label{H_def}
H (\Lambda) = K -
\sqrt{K^2 + \epsilon^2 T_1 T_2 \left[ \Lambda_m^2 - 
(\Lambda - \Lambda_m)^2 \right]}
\end{equation}
with 
\begin{equation}
\Lambda_m =\frac{(\eta_C - \bar{\eta})}{2 T_2} \geq 0.
\end{equation}
Here, $\bar\eta = \langle W\rangle / \langle Q\rangle = 
1-\delta/\epsilon$ is the average efficiency derived in the previous section
and $\eta_C = 1-T_2/T_1$ is the Carnot efficiency.
The derivation and the exact expression for $\mathsf{J}$ are presented in
Appendix~\ref{sec:app}.

After integrating $G_t(x_1,x_2,\lambda_Q,\lambda_W)$ over
$\bmx$, one obtains the reduced generating function 
$\tilde{G}_t(\lambda_Q,\lambda_W)$ for $Q$ and $W$. 
The integration does not introduce an additional $t$-dependent
term in the exponent as far as $\mathsf{J}$ is positive-definite.
Therefore, the cumulant
generating function~(CGF) $\phi(\lambda_Q,\lambda_W) = \lim_{t\to\infty}
\frac{1}{t}\ln \tilde{G}_t(\lambda_Q,\lambda_W)$~\cite{Verley:2014hq} 
is given by
\begin{equation}\label{phi_res}
\phi(\lambda_Q,\lambda_W) = \mu(\lambda_Q,\lambda_W)
\chi_\mathsf{J}(\lambda_Q,\lambda_W), 
\end{equation}
where the characteristic function $\chi_\mathsf{J}(\lambda_Q,\lambda_W)$ 
is equal to unity if the matrix $\mathsf{J}(\lambda_Q,\lambda_W)$ is 
positive-definite and infinity otherwise. 

The LDF $L(\eta)$ is then obtained by using the relation 
\begin{equation}\label{Leta_formula}
L(\eta) = -\min_{\lambda} \phi(-\eta \lambda,\lambda)
\end{equation}
that was derived in Ref.~\cite{Verley:2014hq}. 
To a given value of $\eta$, one need to evaluate the minimum value of the
function $\phi(\lambda_Q,\lambda_W)$ along a straight line $l_\eta$
of slope $-\eta$ passing through the origin
in the $\bm{\lambda} = (\lambda_W,\lambda_Q)$ plane.
Such a task is achieved by using the property of the function $\mu$.
Recall that $\mu(\lambda_Q,\lambda_W) = H(\lambda_Q+\bar\eta \lambda_W)$ 
depends on a single parameter $\Lambda = \lambda_Q+\bar{\eta} \lambda_W$.
Thus, it is constant along a straight line of slope $-\bar\eta$ in the
$\bm{\lambda}$ plane.
The function $H(\Lambda)$ has the minimum value 
\begin{equation}
\mu_m = 
K - \sqrt{K^2 + \epsilon^2 T_1 T_2 \Lambda_m^2} \leq 0
\end{equation}
at $\Lambda = \Lambda_m$, and increases monotonically as $\Lambda$ deviates
from $\Lambda_m$. Thus, $L(\eta)$ is determined by the distance of the line
$l_\eta$ and $\Lambda=\Lambda_m$ inside the domain of $\chi_{\mathsf{J}}=1$.

\begin{figure}
\includegraphics*[width=\columnwidth]{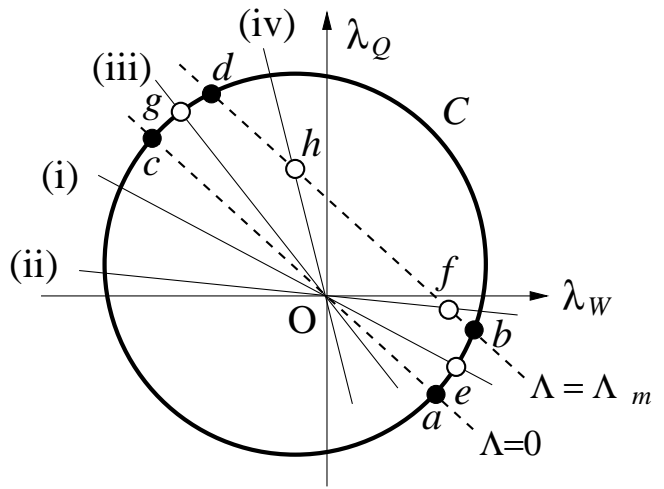}
\caption{ 
In the $\bm\lambda=(\lambda_W,\lambda_Q)$ plane, we draw schematically 
the  boundary $C$ of the $(\chi_{\mathsf{J}}=1)$ domain. Two 
dashed straight lines of slope $-\bar\eta$ correspond to $\Lambda = 0$ 
and $\Lambda_m$ where $\Lambda = \lambda_Q + \bar\eta \lambda_W$.
The intersections between them are marked with closed circles and labeled as
$a$, $b$, $c$, and $d$. 
To a given value of $\eta$, $L(\eta)$ is obtained from the minimum value 
of $\mu(\lambda_Q,\lambda_W)$ along the segment of the straight line 
$l_\eta$ passing through the origin $O$ with slope $-\eta$ 
inside the boundary $C$.
When $\eta = \bar\eta$, the line $l_\eta$ coincides with 
the line $\Lambda=0$ where $\mu=0$. Hence, $L(\bar\eta)=0$.
(i) When $\eta_L \leq \eta < \bar\eta$, the right intersection point $e$ (open
symbol) of $l_\eta$ and ${C}$ lies on a segment between $a$ and $b$. Thus,
the LDF is determined by the $\Lambda$ value at $e$, $L(\eta)
= -H(\Lambda_e)$. The left intersection point is irrelevant since it
is farther from the line $\Lambda=\Lambda_m$ than $e$. 
The territory $\eta_L$ is determined by the condition that the point $e$ 
coincides with $b$.
(ii) When $\eta<\eta_L$, the line $l_\eta$ intersects with the line
$\Lambda=\Lambda_m$ within ${C}$ at point $f$ (open symbol). 
Hence, $L(\eta) = -\mu_m$.
(iii) When $\bar\eta<\eta \leq \eta_R$, the left intersection point 
$g$~(open symbol) lies on a segment between $c$ and $d$.  
Thus, the LDF is given by $L(\eta) = -H(\Lambda_g)$.
The territory $\eta_R$ is determined by the condition that $g$ coincides with
$d$. (iv) When $\eta > \eta_R$, the line $l_\eta$ intersects with the line
$\Lambda=\Lambda_m$ at point $h$~(open symbol), and $L(\eta) = -\mu_m$.
}\label{fig4}
\end{figure}

\begin{figure}
\includegraphics*[width=\columnwidth]{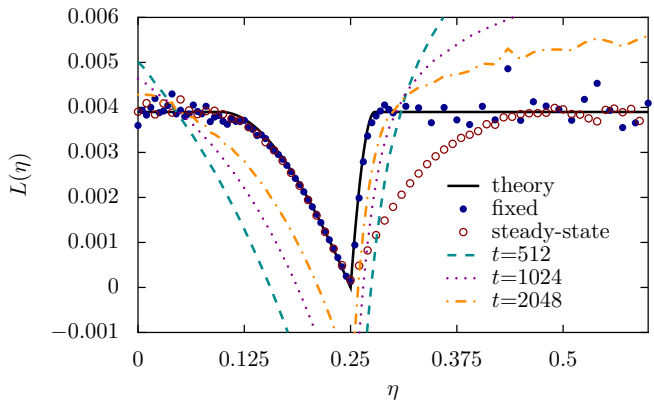}
\caption{
The LDF for efficiency with the set of parameters
$K=1$, $T_1=2$, $T_2=1$, $\epsilon=1/2$, and $\delta=3/8$.
Black line represents the analytic result obtained from
Eq.~(\ref{Leta_formula}).
Filled circles represent the numerical results from the fixed
initial condition, while open circles represent those from the
steady-state initial condition.
The colored lines represent $-\frac{1}{t}\ln P_t(\eta)$
at $t=512$, $1024$, and $2048$.
Details for the simulations are explained in the main text.
}
\label{fig5}
\end{figure}

In Fig.~\ref{fig4}, we explain a graphical method to construct 
$L(\eta)$. This method gives an information on the shape of $L(\eta)$:
$L(\eta=\bar\eta)=0$, $L(\eta)$ increases monotonically as $\eta$ deviates 
from $\bar\eta$ in the region $\eta_L < \eta < \eta_R$, and remains constant
$L(\eta) = -\mu_m \geq 0 $ elsewhere. The boundaries $\eta_L (\leq \bar\eta)$
and $\eta_R ( \geq \bar\eta)$ vary with model parameters. 
In Fig.~\ref{fig5}, we show the plot of $L(\eta)$ obtained from the
analytic method using the parameters 
$K=1$, $T_1=2$, $T_2=1$, $\epsilon=1/2$, and $\delta=3/8$ with the mean
efficiency $\bar\eta = 1-\delta/\epsilon = 1/4$. The LDF takes the minimum
value $0$ at $\eta=\bar\eta$ and a constant value $-\mu_m = (\sqrt{17}-4)/4$
in the regions with $\eta \leq \eta_L \simeq 0.098$ and $\eta\geq \eta_R
\simeq 0.278$.  

We also performed numerical simulations to confirm the analytic result.
Starting from the fixed initial configuration $x_1=x_2=0$, we integrated the 
time-discretized overdamped Langevin equation numerically 
by using the Heun method~\cite{Greiner:1988cqa} with $\Delta t = 0.01$. 
We measured the work and the heat up to time $t=512$, $1024$, and $2048$, and
constructed the probability distribution for the efficiency $P_t(\eta)$ by
repeating the simulations for $N_s = 3\times2^{25} \approx 10^8$ times.
The LDF $L(\eta)$ can be estimated by fitting 
$-\frac{1}{t}\ln P_t(\eta)$ to the function $A(\eta)/t + B(\eta) \ln t / t + 
L(\eta)$ at each value of $\eta$~\cite{Proesmans:2015gt,Proesmans:2015kw}.
The LDF thus obtained is presented in Fig.~\ref{fig5}. The
numerical result is in good agreement with the analytic result: 
$L(\eta)$ is minimum at $\eta=\bar\eta$ and monotonically
increases as $\eta$ deviates from $\bar\eta$ to reach its maximum although
statistical uncertainty becomes noticeable for large $|\eta-\bar\eta|$.

Also shown in Fig.~\ref{fig5} is the LDF obtained from the steady-state
initial condition where $x_1$ and $x_2$ at $t=0$ are drawn from the 
steady-state distribution. Interestingly, the LDF is different from the LDF
obtained from the fixed initial condition. 
The LDF of nonequilibrium fluctuations may be affected by the initial
condition due to the everlasting initial memory effect~\cite{Lee:2013kz}. 
Our results exemplify the initial condition dependent behavior of 
the LDF~(see also Appendix~\ref{sec:app}).

The LDF of our model system does not follow the universal behavior,
suggested in Ref.~\cite{Verley:2014ha,Verley:2014hq},
that the Carnot efficiency is the sole maximum point of the LDF.
We explain the reason for this discrepancy.
In Ref.~\cite{Verley:2014ha,Verley:2014hq},
the least likeliness of the Carnot efficiency was shown for systems
possessing a finite number $\Omega_{sys}$ of microscopic states.
The finiteness of $\Omega_{sys}$ plays a crucial role. The total entropy
production of the engine and two heat baths are given by
$\Delta S_{tot} = - \frac{Q_1}{T_1} - \frac{Q_2}{T_2} + \Delta S_{sys}$ with
the Shannon entropy change $\Delta S_{sys}$ of the system. The energy
conservation requires that $\Delta E = Q_1 + Q_2 - W$ where $\Delta E$
denotes the change in the internal energy of the engine. 
Eliminating $Q_2$, the total entropy production is given by $\Delta
S_{tot} = \frac{\eta_C}{T_2} Q_1 - \frac{1}{T_2}W + ( -\frac{1}{T_2}
\Delta E + \Delta S_{sys})$. Note that the mean value of $Q_1$ and $W$
increases linearly in $t$. On the other hand, when $\Omega_{sys}$ is finite, 
$|\Delta E|$ is bounded above by $(E_{max}-E_{min})$ with the maximum~(minimum)
energy among the $\Omega_{sys}$ states and $|\Delta S_{sys}|$ by $\ln
\Omega_{sys}$.
Consequently, $\Delta S_{sys}$ may be approximated as $\Delta
S_{tot} \simeq \frac{\eta_C}{T_2} Q_1 - \frac{1}{T_2}W$ in the large $t$ 
limit.
When the total entropy production is written as the sum of thermodynamic
quantities, the joint probability distribution of them satisfies the
fluctuation theorem~\cite{GarciaGarcia:2012vl, GarciaGarcia:2010tj, Noh:2012hg,
Noh:2014if} 
\begin{equation}\label{ft_j}
\frac{P(Q_1,W)}{P(-Q_1,-W)} \simeq e^{-\Delta S_{tot}} .
\end{equation}
The least likeliness of the Carnot efficiency is the direct
consequence of the fluctuation theorem~\cite{Verley:2014ha,Verley:2014hq}.

In contrast to the underlying assumption of
Refs.~\cite{Verley:2014ha,Verley:2014hq}, our model has the continuous phase
space and the internal energy of the engine is unbounded. 
Although the averages of $\Delta S_{sys}$ and energy $\Delta E$ 
are zero in the steady state, stochastic fluctuations may generate 
rare events accompanied by $\Delta S_{sys}$ and $\Delta E$ comparable with
$Q_i$ and $W$~\cite{Nemoto:2012kp}. 
It is known that such fluctuations are nonnegligible and 
invalidate the fluctuation theorem derived by ignoring
them~\cite{{VanZon:2003wh},Noh:2012hg,Noh:2014if}.
Therefore, we conclude that the least likeliness of the Carnot efficiency 
based on the fluctuation theorem in \eqref{ft_j} is not valid in our model.


\section{Summary and discussions}\label{sec:Conclusion}
In this paper, we introduced a model for a heat engine which operates 
between two heat baths and is driven by a nonconservative force. 
The model is described by
a Ornstein-Uhlenbeck process and most of the properties are analytically
tractable. Firstly, we showed that the efficiency at maximum power is given
by the so-called Curzon-Ahlborn efficiency $\eta_{MP} = \eta_{CA} =
1-\sqrt{T_2/T_1}$. This is a surprising result because 
$\eta_{CA}$ has been believed to be the property of the endoreversible engine 
while our engine is not endoreversible. Instead, we showed that $\eta_{CA}$
is the consequence of the relation $s_1 = \mathcal{F}(s_2)$ between the
entropy loss $s_1$ of the hot bath and the entropy gain $s_2$ of the cold
bath with the universal function given in~\eqref{eq:f(x)}. 

Secondly, we derived the analytic expression for the LDF $L(\eta)$ 
of the efficiency fluctuation. The shape of $L(\eta)$ is shown in
Fig.~\ref{fig5}: It is minimum at $\eta = \bar\eta$ and displays plateaus
far from $\bar\eta$. Our result shows that $L(\eta)$ does not have 
a peak at the Carnot efficiency $\eta_C$. 
Thus, the least likeliness of the Carnot efficiency is
limited to systems only with a finite number of microscopic states. 
Nonequilibrium fluctuations in systems with continuous degrees of freedom
invalidate the least likeliness of $\eta_C$. We also found that the LDF of
the efficiency depends on the initial condition, which stresses 
the initial memory effect of nonequilibrium systems~\cite{Lee:2013kz}.

The linear solvable model has provided a lot of informations on the
properties of the heat engines. It also suggests interesting theoretical 
questions. 
It is shown that the the Curzon-Ahlborn efficiency at maximum power is
guaranteed by the relation $s_1 = \mathcal{F}(s_2)$ with the universal
function $\mathcal{F}(x)$ given in \eqref{eq:f(x)}.
On the other hand, the Curzon-Ahlborn efficiency was investigated from 
the viewpoint of symmetry in 
Refs.~\cite{{Van-den-Broeck:2005aa},{Esposito:2009ke},{Cleuren:2015aa}}.
It would be interesting to pursue the implication of the relation $s_1 =
\mathcal{F}(s_2)$ on underlying symmetry of engine dynamics. The LDF
$L(\eta)$ for the system under the steady-state initial condition requires
the whole eigenstates of the Fokker-Planck operator, which are not available
yet. We would like to leave those tasks for future works.

\begin{acknowledgments}
This research was supported by the National Research Foundation (NRF) of
Korea Grant No. 2013R1A2A2A05006776. 
\end{acknowledgments}

\appendix

\section{Derivation of $L(\eta)$ and discussion on the initial condition 
dependency}\label{sec:app}

In the overdamped limit,
the infinitesimal heat and work in \eqref{eq:HeatWork} 
during the time interval $dt$ are given by 
$\dbar Q_1 = -f_1 \circ dx_1 = -f_1^2 dt - f_1 \circ d\Xi_1$ and 
$\dbar W = -\bmf_{nc} \circ d\bmx = -(\bmf_{nc} \cdot \bmf) dt -f_{nc,1} \circ
d\Xi_1 - f_{nc,2}\circ d\Xi_2$~($\gamma$ is set to 1).
We will use $\cdot$ for the inner product of a vector with another vector or
a matrix.
Hence, $\bmy = (x_1,x_2,Q=Q_1,W)^T$ follows a stochastic differential
equation 
\begin{equation}
\dot{\bmy} = \bm{d} + \mathsf{N} \cdot \bmzeta ,
\end{equation}
where the drift vector $\bm{d} = (f_1,f_2,-f_1^2,-\bmf \cdot \bmf_{nc})^T$, 
the $(4\times 2)$ noise matrix
$\mathsf{N}$ is given by
\begin{equation}
\mathsf{N} = \begin{pmatrix}
  \sqrt{2T_1} & 0 \\
  0 & \sqrt{2 T_2} \\
  - \sqrt{2T_1} f_1 & 0 \\
  - \sqrt{2T_1} f_{nc,1} & -\sqrt{2T_2} f_{nc,2}
\end{pmatrix},
\end{equation}
and the components of the noise vector $\bm{\zeta}(t)=
(\zeta_1(t),\zeta_2(t))^T$
are independent Gaussian random variables of zero mean and unit variance.
The total force $\bmf = \bmf_{c} + \bmf_{nc}$ is linear in $\bmx$, hence it
is written as $\bmf = -\mathsf{F} \cdot \bmx$ with the force matrix
$\mathsf{F}$.

The differential equation is nonlinear and involves the multiplicative
noises
implemented with the Stratonovich interpretation. Following the standard
recipe~\cite{Risken:1996vl}, one can derive the Fokker-Planck equation for
the probability distribution $p(\bmy;t)$:
\begin{equation}
\frac{\partial p}{\partial t} = \mathcal{L} p
\end{equation}
where the Fokker-Planck operator is given by
\begin{equation}\label{FPO}
\mathcal{L} = - \bm{\nabla}^T \cdot \bm{d} + \frac{1}{2} (\bm\nabla^T \cdot
\mathsf{N}) \cdot (\bm\nabla^T \cdot \mathsf{N})^T
\end{equation}
with the differential operator $\bm\nabla =
\left(\frac{\partial}{\partial x_1},\frac{\partial}{\partial
x_2},\frac{\partial}{\partial Q},\frac{\partial}{\partial W}\right)^T$.

For $G(x_1,x_2,\lambda_Q,\lambda_W;t)$ defined in \eqref{G_gf},
the time evolution operator ${\mathcal{L}}_\lambda$ 
is obtained by 
replacing $\partial/\partial Q$ and
$\partial/\partial W$ in $\mathcal{L}$ with $\lambda_Q$ and $\lambda_W$,
respectively. The resulting operator becomes bilinear in $\bmx$ and
the gradient operator 
$\bm\nabla_\bmx = \left(\frac{\partial}{\partial x_1},
\frac{\partial}{\partial x_2}\right)^T$, i.e.,
\begin{equation}\label{L_lambda}
{\mathcal{L}}_\lambda =
\bm\nabla_\bmx^T \cdot \mathsf{D} \cdot \bm\nabla_\bmx
+ 2 \bmx^T \cdot \mathsf{B}^T \cdot \bm\nabla_\bmx
+ \bmx^T \cdot \mathsf{A} \cdot \bmx
+ K + \rm{Tr} \mathsf{B},
\end{equation}
where $\mathsf{A}$ and $\mathsf{D} = {\rm diag}(T_1,T_2)$ are the 
$2\times 2$ symmetric matrices,
$\mathsf{B}$ is the $2\times 2$ nonsymmetric matrix, and $\rm{Tr}\mathsf{X}$
denotes the trace of a matrix $\mathsf{X}$.
The matrix elements for $\mathsf{A}$ and $\mathsf{B}$ are readily read from
\eqref{FPO}. Explicitly, they are given by $\mathsf{A} = \mathsf{C}^T
\mathsf{D} \mathsf{C} + \frac{1}{2} \mathsf{F}^T \mathsf{C} + \frac{1}{2}
\mathsf{C}^T \mathsf{F}$ and $\mathsf{B} = \mathsf{D C}
+\frac{1}{2}\mathsf{F}$ with an auxiliary matrix
\begin{equation}
\mathsf{C} = 
\begin{pmatrix} 
K \lambda_Q & -\epsilon (\lambda_Q + \lambda_W) \\
-\delta \lambda_W & 0  
\end{pmatrix} .
\end{equation}

We now rescale the coordinates to define $\bmz = (\hat{x}_1,\hat{x}_2)^T 
\equiv \mathsf{D}^{-1/2} \cdot \bmx$. Then,
the time evolution operator is rewritten in terms of $\bmz$ as
\begin{equation}
{\mathcal{L}}_\lambda =
\nabla_\bmz^2
+ 2 \bmz^T \cdot \hat{\mathsf{B}}^T \cdot \bm\nabla_\bmz
+ \bmz^T \cdot \mathsf{D} \hat{\mathsf{A}} \cdot \bmz
+ K + \rm{Tr} \mathsf{B}
\end{equation}
where $\bm\nabla_\bmz = \left(\frac{\partial}{\partial \hat{x}_1},
\frac{\partial}{\partial \hat{x}_2}\right)^T$
and $\hat{\mathsf{X}}=\mathsf{D}^{-1/2}\cdot\mathsf{X}\cdot\mathsf{D}^{1/2}$
for any matrix $\mathsf{X}$.
It looks similar to the Hamiltonian of the two-dimensional harmonic
oscillator except for the second term.
Finally, we make a transformation 
\begin{equation}
\widetilde{\mathcal L}_\lambda \equiv e^{\frac{1}{2} \bmz^T \cdot \mathsf{J}
\cdot \bmz} {\mathcal L}_\lambda e^{-\frac{1}{2} \bmz^T \cdot \mathsf{J}
\cdot \bmz}
\end{equation}
with a certain symmetric matrix $\mathsf{J}$ which will be determined later.
It acts as the time evolution operator for the modified generating function
$e^{\frac{1}{2} \bmx^T \cdot \mathsf{D}^{-1/2} \mathsf{J} \mathsf{D}^{-1/2}
\cdot \bmx}
G(\bmx,\lambda_Q,\lambda_W;t)$.
This transformation replaces the gradient operator $\bm\nabla_\bmz$ 
with
$\bm\nabla_\bmz-\mathsf{J}\cdot\bmz$, which leads to
\begin{equation}\label{Ltilde}
\widetilde{\mathcal L}_\lambda =
\nabla_\bmz^2
- 2 \bmz^T \cdot \mathsf{M}^T \cdot \bm\nabla_\bmz
+ \bmz^T \cdot \mathsf{Q} \cdot \bmz
+ \mu,
\end{equation}
where
\begin{equation}
\begin{split}
 \mathsf{M} &= \mathsf{J}-\hat{\mathsf{B}}, \\
 \mathsf{Q} &= \mathsf{M}^T \mathsf{M} + \mathsf{D} \hat{\mathsf{A}} -
 \hat{\mathsf{B}}^T \hat{\mathsf{B}}  
 =\mathsf{M}^T \mathsf{M} -
                \frac{1}{4}\hat{\mathsf{F}}^T\hat{\mathsf{F}},\\
 \mu        &= K-\rm{Tr }~ \mathsf{M}.
\end{split}
\end{equation}

The operator $\widetilde{\mathcal L}_\lambda$ is
simplified if one chooses $\mathsf{J}$ or $\mathsf{M}$ suitably
so that $\mathsf{Q}=0$. 
It is accomplished by choosing 
\begin{equation}\label{J_ab}
\mathsf{J} = \hat{\mathsf{B}} + \frac{1}{2}\mathsf{O}\hat{\mathsf{F}}
\end{equation}
with an orthogonal matrix
\begin{equation}
\mathsf{O} = \begin{pmatrix}
  \cos \theta & -\sin \theta \\
  \sin \theta & \cos \theta
\end{pmatrix}.
\end{equation}
The angle variable $\theta$ has to be determined by requiring that
$\mathsf{J}$ should be a symmetric matrix.
Then, the time evolution operator $\widetilde{\mathcal L}_\lambda$
has the constant eigenfunction with the corresponding eigenvalue $\mu$.
As a result, in the large $t$ limit, 
the generating function $G$ has the asymptotic form in (\ref{Gt_sol}).

We find that the symmetry condition $J_{12} = J_{21}$ is satisfied if 
\begin{equation}\label{theta_ab}
\theta = \alpha \pm \beta
\end{equation}
where
$\cos\alpha = \frac{\hat{F}_{12}-\hat{F}_{21}}{R}$,
$\sin\alpha = -\frac{\hat{F}_{11}+\hat{F}_{22}}{R}$,
$\cos\beta = \frac{2 \left( \hat{B}_{21}-\hat{B}_{12} \right)}{R}$,
and $\sin\beta = \sqrt{1-\cos^2\beta}$
with 
\begin{equation}
R = \sqrt{\left( \hat{F}_{12}-\hat{F}_{21} \right)^2 + \left(
\hat{F}_{11}+\hat{F}_{22} \right)^2}.
\end{equation}
There are two different solutions for $\mathsf{J}$
due to the sign ambiguity in (\ref{theta_ab}).
To select the proper solution, we impose the condition that
the generating function $G \left( \bmx, \lambda_Q, \lambda_W ; t \right)$
in the infinite $t$ limit 
should converge to the steady-state distribution when
$\lambda_Q = \lambda_W = 0$.
The steady-state probability distribution of a linear system is known
exactly~\cite{Kwon:2011fk}.
Comparing the two solutions with the steady-state probability 
distribution, we find that the matrix $\mathsf{J}$ is indeed given by
$\mathsf{J} = \hat{\mathsf{B}}+\frac{1}{2} \mathsf{O} \hat{\mathsf{F}}$
with $\theta = \alpha + \beta$. The eigenvalue is given by
\begin{equation}
\mu \left( \lambda_Q, \lambda_W \right)
= K - \frac{R}{2} \sin\beta \ ,
\end{equation}
which yields the result in (\ref{mu_res}).

We add a remark on the initial condition dependence. 
In our calculation, we keep only the leading eigenvalue $\mu$ of 
$\widetilde{\mathcal{L}}_\lambda$. Integration over the final position
$\bmx$ introduces a cut represented by the characteristic function 
$\chi_\mathsf{J}$ to yield the result in \eqref{phi_res}. 
As for the effect of the initial condition, Visco studied a similar 
problem, a Brownian particle in one dimension in contact with 
two heat baths~\cite{Visco:2006en}. Visco obtained the
exact moment generating functions for both the fixed and the steady-state 
initial conditions by considering all the eigenstates of the Fokker-Planck 
operator. The study reveals that fluctuations in the initial configuration
introduce an additional characteristic function.
In this regard, 
we expect that the steady-state initial condition in our model modifies the
characteristic function $\chi_\mathsf{J}$ 
so that the LDF $L(\eta)$ in the steady-state initial condition is broader.
We confirm this expectation with numerical simulations. In
Fig.~\ref{fig5}, we compare the LDFs from the fixed initial condition and
from the steady-state initial condition. One finds that the LDF from the 
latter displays a broader distribution in the $\eta \geq \bar\eta$ side
although the analytic expression for that is not available yet.

\bibliographystyle{apsrev}
\bibliography{paper}

\end{document}